\documentclass[12pt]{iopart}

\usepackage{iopams}

\usepackage[dvips]{graphicx}

\begin{document}

\title{Solvable Metric Growing Networks}

\author{M O Hase and J F F Mendes}

\address{Departamento de F\'isica da Universidade de Aveiro, 3810-193 Aveiro, Portugal}
\ead{mhase@if.usp.br}

\begin{abstract}
Structure and dynamics of complex networks usually deal with degree distributions, clustering, shortest path lengths and other graph properties. Although these concepts have been analysed for graphs on abstract spaces, many networks happen to be embedded in a metric arrangement, where the geographic distance between vertices plays a crucial role. The present work proposes a model for growing network that takes into account the geographic distance between vertices: the probability that they are connected is higher if they are located nearer than farther. In this framework, the mean degree of vertices, degree distribution and shortest path length between two randomly chosen vertices are analysed.
\end{abstract}


\maketitle


\section{Introduction}
\label{introduction}

For some decades the paradigm of network was represented by the random graphs\cite{B85}, which modelled graphs with a huge number of vertices and edges as a stochastic process. In the well - known Erd\"os - R\'enyi model with $N$ vertices, the occurrence of an edge, from the ${N \choose 2}$ possibilities, is determined independently with some fixed probability. Despite the fact that the introduction of probability concepts in graph theory represented a progress in many areas -- even besides mathematics --, it failed to describe many properties of ``real networks'' like Internet, air traffic system, social links between people, and so on. However, the 1990s witnessed many important progress in attempt to examine these ``real networks''. To cite some of them, the small - world network of Watts and Strogatz\cite{WS98} (see also \cite{NW99}) is a model that can interpolate, by tuning a single parameter, a random graph and a regular lattice. This model succeeded in characterize an interval for this parameter where the small - world property is displayed. Barab\'asi and Albert, on the other hand, have proposed the growing network with preferential linking\cite{BA99}, which succeeded in display the scale - free properties, present in a myriad variety of ``real networks''. These models have been intensively analysed, with many progress.

Nevertheless, the networks studied by the above authors lied on an abstract space, where the geometric structure on which the vertices are placed is not important. This picture revealed to be favourable for networks that can be analysed, for instance, by tools from spin glass theory or spin systems with finite connectivity that can be studied with similar techniques\cite{VB85, MM04, ZM06, HM08}. Many ``real networks'', however, have an additional feature: they are embedded in some geographic structure, which not only allows one to distinguish ``near'' and ``far'' vertices from some point, but they may determine the dynamics of the evolution of networks. For instance, many cities grow starting from a ``initial'' region and expand away from this center gradually; as another example, it is common to see people making friendships with those that live nearer than farther. This theme -- the influence of the geography on networks -- has attracted attention, and there are interesting researches being done in the context of computer science\cite{K00, MvN04, BBCdSK04, FKP02}, with recent progress also in the physical literature\cite{SM03, RbA06}.

In the literature, networks on Euclidean structures have usually other properties attached together\cite{YJB02, MS02, XBS02}. For instance, in \cite{MS02}, a random growing network was considered on a two-dimensional Euclidean structure together with preferential linking; the authors showed, by numerical methods, that the strength of the interaction (that depends on the Euclidean distance between them) between vertices can change the profile of the degree distribution from a power-law to a stretched exponential (similar results were found by \cite{XBS02}), and the small-world property is preserved for any strength of this interaction\cite{XBS02}. A related problem was studied in \cite{B03}, where the crossover between scale-free and spatial networks was considered. The authors of \cite{MMK05} have succeeded in constructing a non-growing geographical small-world network with scale-free distribution. There are also works that investigated weight properties (differences in the link ``intensities'')\cite{BBV05, MMK05} on models related to the one studied here.

Nevertheless, it is important to be aware of the real importance of the metric structure on the network. Therefore, a graph which has metric properties only is an important model since it concentrates on the influence of geographic structure on the network and leads to a better notion of what the real range of a metric network is. For instance, scale-free properties is not expected for the model proposed in this work since it is purely metric and no preferential attachment is invoked. Searching some general features of a network on a metric structure is the main purpose of this work.

The present work intends to contribute in this direction by introducing a growing network that lies on a space with a metric arrangement, where geographic distances between vertices are present and do influence on graph properties of the system. The model proposed in this work allows one to define vertices that are ``far'' or ``close'' from a fixed vertex from the standpoint of their geographic arrangement -- like what happens with a Euclidean space like $\mathbb{R}^{2}$. To stress the metric structure of the model, a power - law form, with exponent $\alpha$, was chosen to describe the probability that two vertices, separated by a metric distance $r$, are connected (the probability is proportional to $r^{-\alpha}$). The motivation to choose this particular form is twofold: firstly, the power - law form can tune, with a single paramenter ($\alpha$), the strength of interaction between vertices (low values of $\alpha$ makes the network close to a random graph, while large values of $\alpha$ makes the network close to a regular lattice structure). In the literature, the exponential distribution has already been tested by Waxman\cite{W88} (moreover, the random network topology generator of Waxman yields a graph that is quite different from the one proposed here, where the distance between vertices are ``carefully'' chosen). Moreover, power - law form is interesting since it yielded many non - trivial results to problems that are similar to ones that are discussed in this work\cite{K00, MvN04}. The model proposed here is manageable to analytic treatments in some interesting cases, and an analysis of the mean degree is carried out, as well as other graph properties.

The next two sections present the model and fix notations. The mean degree of the network is analysed in sections IV and V. The aim to discuss this quantity is to show that the model proposed in this work exhibits a metric structure that shares the same notion of distance present in models embedded in a Euclidean space like, say, $\mathbb{R}^{2}$. For sake of completeness and to stress the fact that the model does show metric structure, the sections VI and VII discuss, respectively, the degree distribution and an estimation of the shortest path length between two randomly chosen vertices in the graph. Some final comments are made in the section VIII.


\section{Notations}
\label{notations}

Let $x$ be a continuous variable that tends to infinity or some limiting value and let $g$ be a positive function of $x$ and consider $f$ as any other function (of $x$). Throughout this work, some symbols\cite{H75} will be extensively used, which are:

\medskip
\noindent
I) $\mathcal{O}$: If $f=\mathcal{O}(g)$, then $|f|<Ag$, where $A$ does not depend on $x$, for all values of $x$ in question.

\medskip
\noindent
II) $o$: If $f=o(g)$, then $f/g\rightarrow 0$.

\medskip
\noindent
III) $\ll$ (or $\gg$): $y\ll z$ will always be understood in the sense that $y/z=o(1)$ \textit{if} no special conditions are specified.

\medskip
\noindent
IV) $\mathcal{O}(x,y)$: Indicates $\max\{\mathcal{O}(x),\mathcal{O}(y)\}$ ($x$ and $y$ are assumed to be non - negative).

\bigskip
In this work, a network will be defined on a metric structure, and the distinction between ``(metric) distance'' and ``path length'' should be clarified. The ``(metric) distance between two vertices, $x_{1}$ and $x_{2}$, is the separation between them on the metric structure, and no reference is made to the links between vertices. On the other hand, the ``path length'' between $x_{1}$ and $x_{2}$ is the minimum number of links that joins them. In the example given below (Figure 1), the (metric) distance between the vertices $x_{a}$ and $x_{b}$ is $5$, but the path length connecting them is $2$.

\begin{figure}[htb]
\centering
\includegraphics[scale=0.6]{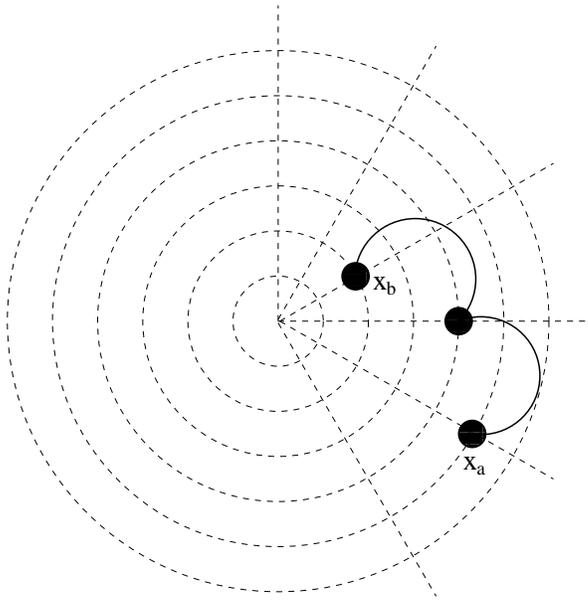}
\caption{Example: the metric distance between each adjacent intersection of dashed lines is taken to be $1$. The bold curves are links between vertices.}
\label{1}
\end{figure}


\section{Model}
\label{model}

Let $0$ be an initial vertex (the ``central node'') from which the network grows toward $m$ branches. At each time step, taken as unity, $m$ new vertices are added to the network, one for each branch. Each one of the new vertex (there are $m$ of them: one for each branch) is added at distance $1$ from the previous vertex in the same branch. Moreover, the distance between two vertices located at adjacent branches will be taken as being $1$ if they were born at the same time (this situation is valid for all $m$, except when $m=1$). Choose now one branch and denote it by $1$; the branches $2$, $3$, $\cdots$, $m$ are counted in counter clockwise orientation, as shown in Figure 2.

\begin{figure}[htb]
\centering
\includegraphics[scale=0.6]{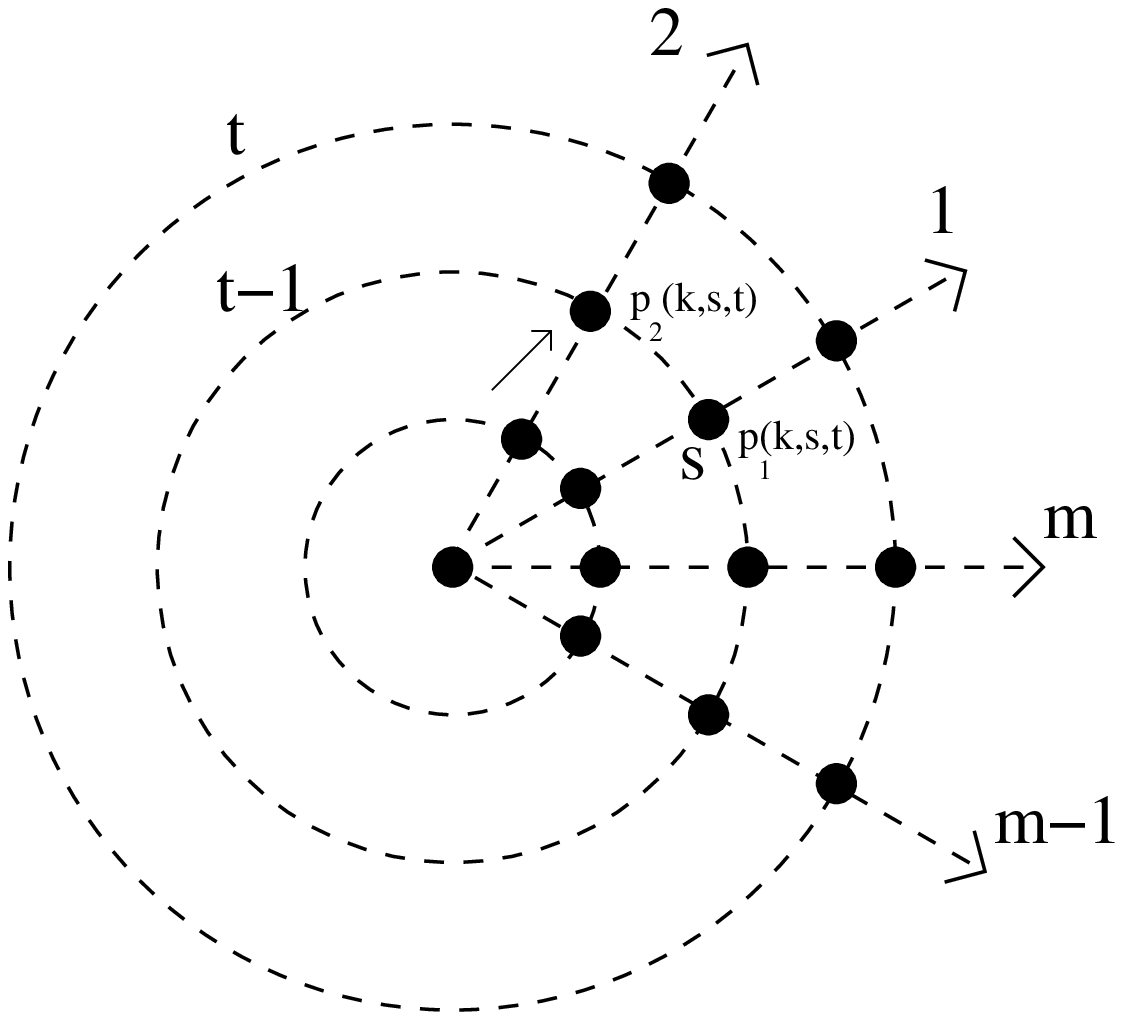}
\caption{Growing network with $m$ branches at time $t$. The probability that the vertex born at time $s$ (located at branch $1$) has $k$ links is $p_{1}(k,s,t)$. Note that in this example, $t=3$ and $s=t-1=2$. By the symmetry of the problem, the probability $p_{2}(k,s,t)$ that a vertex located at the branch $2$ (this vertex is indicated by an arrow in the picture) has $k$ links should be the same: $p_{1}(k,s,t)=p_{2}(k,s,t)$. Following the same argument, one has $p_{1}(k,s,t)=p_{2}(k,s,t)=\cdots=p_{m}(k,s,t)$.}
\label{2}
\end{figure}

In this work, the probability $p_{i}(k,s,t)$ that a vertex, which was born at time $s$ ($\leq t$) at branch $i$, has $k$ links at time $t$ plays the major role. However, the symmetry of the model shows that the $m$ functions $p_{1}(k,s,t)$, $p_{2}(k,s,t)$, $\cdots$ $p_{m}(k,s,t)$ are equal. Thus, for simplicity of notation, this work will deal with just one of them, say $p_{1}(k,s,t)$. Now denote $p_{1}(k,s,t)$ by $p(k,s,t)$. The function $p(k,s,t)$ corresponds to a particular choice of one vertex (born at time $s$) among $m$ without losing generality. This selected vertex will be considered to be located, by notation, at the ``branch $1$''. This choice does not affect the analysis, since the only important quantity involving branches are their differences.

After $t$ ($>1$) time steps, a vertex born at time $s$, with $0<s<t$, will have four nearest neighbors which are at distance $1$ (except when $m=1$ or $m=2$): the vertices $s-1$ and $s+1$ (both of them belonging to the same branch), and the vertices that were born at the same time at the (two) adjacent branches. This sets the positions that vertices will occupy in the network, but it does not say how they link to each other, which will be described below.

Summarizing the model presented above, one has the following notations:

\medskip

a) $p(k,s,t)$ denotes the probability that a vertex born at time $s$ has $k$ links at time $t$. This fixes a specific vertex (denoted as $s$) among $m$ vertices that have also born at the same time.

b) This \textit{fixed} vertex $s$ will be considered to belong at branch $1$, from which the branches are counted up to $m$ in counter clockwise orientation.

c) By ``vertex $s$'' one should understand the vertex which was born at time $s$ or, equivalently, a vertex which is at distance $s$ from the central node.

d) The parameter $t$ refers to the ``time'' of the network growing process and also indicates the distance of the ``border'' of the network to the central node.

\medskip

Now consider a growing network with $m$ branches and fix a vertex $s$ (at branch $1$, by convention). At time $t(>s)$, each one of the $m$ new vertices that joins the network is made to connect to \textit{one} of the $m(t-1)+1$ old vertices. The probability that a new vertex (born at time $t$), located at branch $b$, links to the selected vertex $s$ (at branch $1$) is denoted by $w_{b}^{(m)}(s,t)$. This probability is made to decay algebrically with the distance between vertices, and its explicit form is shown during the next sections. The distance is measured according to the previously defined network structure: a vertex $s$ ($0<s<t$) is at distance $1$ from its nearest neighbors along its branch and also at distance $1$ with its nearest neighbors, born at the \textit{same} time $s$, located at adjacent branches; see Figure 2. These latter two neighbors are absent when $m=1$ and $m=2$ (when there are only two vetices that join the network at each time). The power - law form of $w_{b}^{(m)}(s,t)$ enables one to observe the effects of the range of the interaction between vertices, as can be seen below.

In this work, rewiring and decimation of edges are not allowed.


\section{One branch}
\label{one_branch}

Consider a network which grows along one direction only (as a ``queue''), which is the case $m=1$. Each new vertex, $t+1$ (henceforth, by convenience, new vertices will born at time $t+1$ instead of $t$), is linked to just one old vertex, say $s$, with probability $w_{1}^{(1)}(s,t+1)$, which depends on the distance (along the line the vertices are placed) between them. For simplicity, $w_{1}^{(1)}(s,t+1)$ will be denoted as $w(s,t+1)$ in this one branch case; the notation $w_{b}^{(m)}$ will be useful later, in the $m$ branches case.

\begin{figure}[htb]
\centering
\includegraphics[scale=0.6]{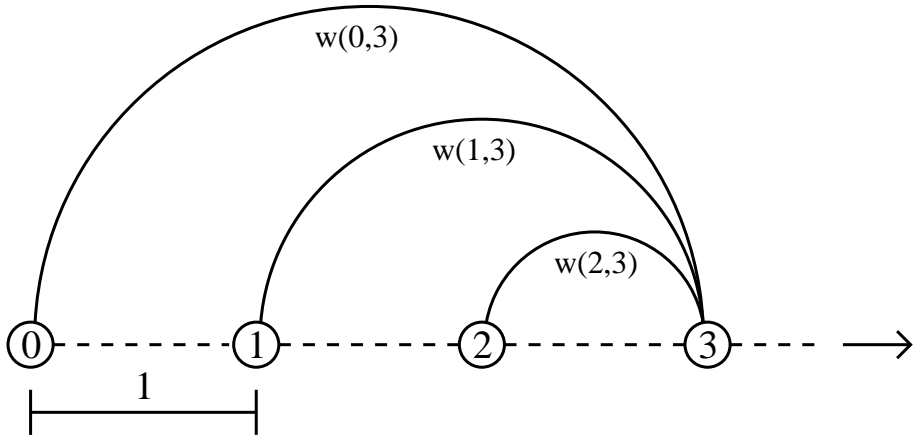}
\caption{Rule for the link of the $(t+1)$ born vertex (here, $t+1=3$). The vertex $3$ links vertex $0$, $1$ and $2$ with probability $w(0,3)$, $w(1,3)$ and $w(2,3)$, respectively.}
\label{se}
\end{figure}



Denoting by $p(k,s,t)$ the probability that a vertex $s$ has $k$ links at time $t$, the master equation of the growing process of this network is casted as

\begin{eqnarray}
p(k,s,t+1) = w(s,t+1)p(k-1,s,t) + \overline{w}(s,t+1)p(k,s,t)\,,
\label{me1}
\end{eqnarray}
where $\overline{w}(s,t):=1-w(s,t)$, and is subjected to the condition $p(k,s=t,t>0)=\delta_{k,1}$, which means that a new vertex has, initially, one (``own'') link.

At time $t+1$, the probability $w(s,t+1)$ of a vertex $s$ be connected by the new incoming vertex $t+1$ is
\begin{eqnarray}
w(s,t+1) = \left\{
\begin{array}{lcl}
\displaystyle\frac{\left(t+1-s\right)^{-\alpha}}{\sum_{r=1}^{t}r^{-\alpha}} & , & s<t+1 \\
 & & \\
0                                                                           & , & s\geq t+1
\end{array}
\right.\,.
\label{w1}
\end{eqnarray}
In particular, $w(s=t,t)=0$ excludes the formation of ``tadpoles'' (a vertex that links itself). In the one branch case, both the $s$ vertex and the new born $t+1$ vertex belong always to the same branch. This may not be the case when more branches are present.

Defining
\begin{eqnarray}
\langle k\rangle(s,t) := \sum_{k=1}^{\infty} kp(k,s,t)
\label{avk}
\end{eqnarray}
as the mean degree of the vertex $s$ at time $t$, its equation of moviment can be casted as
\begin{eqnarray}
\langle k\rangle(s,t+1) - \langle k\rangle(s,t) = w(s,t+1)
\label{movk1}
\end{eqnarray}
by using the master equation (\ref{me1}).

Taking into account the condition $\langle k\rangle(s=t,t)=1$ (which is derived from $p(k,s=t,t>0)=\delta_{k,1}$), the solution of this equation can be determined as
\begin{eqnarray}
\langle k\rangle(s,t) = \left\{
\begin{array}{lcl}
1 + \displaystyle\sum_{r=s+1}^{t}w(s,r) & , & s<t \\
 & & \\
1 & , & s=t
\end{array}
\right.
\,.
\label{k1}
\end{eqnarray}

The behaviour of $\langle k\rangle(s,t)$ depends on the order of magnitude of both $s$ and $t$. The analysis will focus on the regime $1\ll s$, which may, at first, seem to ignore the region of the network ``close'' to the central node; however, for sufficiently large $t$ (in other words, for sufficiently large network), the \textit{relative} position $s$, initially chosen to be farther from the point $0$, is translated to a region near to the central node. Therefore, for the purposes of this work, selecting $s$ to be much larger than $1$ is not a serious restriction. As a remark, the regime $1\ll s\ll t$ enables one to cast the continuous version of the equation (\ref{movk1}), although this procedure of transforming a difference equation to a differential equation is not necessary, since the solution (\ref{k1}) could be obtained.

Before presenting the results, equipping this $m=1$ model with preferential linking term (which makes highly connected vertices more likely to receive more links), leads to an already studied work\cite{DM00}.

In the regime $1\ll s\ll t$, which leads the relative position $s$ on the network closer to the central node for sufficiently large $t$, one has
\begin{eqnarray}
\langle k\rangle(s,t) = \left\{
\begin{array}{lcl}
\left(1-\alpha\right)\displaystyle\ln\left(\frac{t}{s}\right) + \mathcal{O}(1) & , & 0\leq\alpha<1 \\
 & & \\
\displaystyle\ln\left(\frac{\ln t}{\ln s}\right)\left[1+o(1)\right] & , & \alpha=1 \\
 & & \\
2 + \mathcal{O}(s^{1-\alpha}, \tau^{1-\alpha}) & , & \alpha>1
\end{array}
\right.\,.
\label{1_ll_s_ll_t}
\end{eqnarray}
The mean degree in a more general regime, $1\ll s<t$, is presented in the Appendix \ref{onebranchcase}. For $\alpha\leq 1$, the mean degree is always increasing with the size of the network. Nevertheless, when $\alpha$ becomes larger than $1$, the contribution to the degree of a vertex comes mainly due to the next vertex only. The calculations have assumed that $s\ll t$ in the sense that $\ln(t/s)\gg 1$ and $\ln(\ln t/\ln s)\gg 1$ for $0\leq\alpha<1$ and $\alpha=1$, respectively (theses points are detailed in the Appendix A.1).


\section{$m$ branches}
\label{m_branches}

The general case, where the network grows toward $m$ ($>1$) directions, shows that the number of branches plays a major role in the behaviour of the mean degree of a vertex.

The probability that a new born vertex, at branch $b$ (and time $t+1$), links to an old one, born at time $s$ (and located, by convention, at branch $1$), is given by
\begin{eqnarray}
w_{b}^{(m)}(s,t+1) = \frac{d^{(m)}(s,t+1,b)^{-\alpha}}{N(t+1,m,\alpha)}\,,
\label{w}
\end{eqnarray}
where
\begin{eqnarray}
N(t,m,\alpha) := t^{-\alpha} + \sum_{u=1}^{t-1}\sum_{h=1}^{m} d^{(m)}(u,t,h)^{-\alpha}
\label{norm}
\end{eqnarray}
is the normalization (see Appendix \ref{app_normalization}) and
\begin{eqnarray}
\nonumber d^{(m)}(s,t,b) & := & t - s + \min\big\{2s, \min\{b-1, m-b+1\} \big\} \\
\label{dstb}
\end{eqnarray}
is the distance between these two vertices. This is a natural generalization of the previous probability (\ref{w1}). Here, the distance should take into account the fact that the minimum path (over the metric structure) between a new born vertex and some old vertex may cross the central node (this possibility is not seem for $m\leq 5$). This explains the term $\min\big\{2s, \min\{b-1, m-b+1\} \big\}$, where $2s$ stands for a path (over the metric structure) that passes along the central node, and $\min\{b-1, m-b+1\}$ is appliable when the shortest ``metric'' path realizes contourning the central node. For the model with $m$ branches, the master equation is written as
\begin{eqnarray}
\fl
\nonumber p(k,s,t+1) = p(k-m,s,t)\prod_{j=1}^{m}w_{j}^{(m)}(s,t+1) + p(k,s,t)\prod_{j=1}^{m}\overline{w}_{j}^{(m)}(s,t+1) + \\
\fl + \sum_{b=1}^{m-1} p(k-b,s,t)\sum_{q_{1}<\cdots<q_{b}}w_{q_{1}}^{(m)}(s,t+1)\cdots w_{q_{b}}^{(m)}(s,t+1) \prod_{\stackrel{j=1}{j\neq q_{1},\cdots,q_{b}}}^{m} \overline{w}_{j}^{(m)}(s,t+1)\,,
\label{me}
\end{eqnarray}
where, in the right side, the first term represents the case when all the ($m$) new vertices links to a specific one $s$ and the second term is when none of the $m$ new vertices links $s$. The third term is the sum of the remaining intermediate cases.

As in the previous section, one is interested in the mean degree of a vertex $s$ at time $t$, namely $\langle k\rangle(s,t)$. From the definition (\ref{avk}) and the master equation (\ref{me}), it is possible to show, after a lengthy arithmetic manipulation, that (see Appendix \ref{mmeandegree})
\begin{eqnarray}
\langle k\rangle(s,t+1) = \langle k\rangle(s,t) + \sum_{j=1}^{m}w_{j}^{(m)}(s,t+1)\,.
\label{keqm}
\end{eqnarray}

The solution of the difference equation (\ref{keqm}) is
\begin{eqnarray}
\langle k\rangle(s,t) = \left\{
\begin{array}{lcl}
1 + \displaystyle\sum_{b=1}^{m} \sum_{r=s+1}^{t} w_{b}^{(m)}(s,r) & , & s<t \\
 & & \\
1 & , & s=t
\end{array}
\right.\,.
\label{k}
\end{eqnarray}

To stress the influence of the number of branches on the graph property of the network, the mean degree will be calculated in the regimes $1<m\ll s\ll t$, where the number of branches is small, and $1\ll s\ll t\ll m$, where $m$ is large. Actually, in the former case ($1<m\ll s\ll t$), one has the same results for the $m=1$ model:
\begin{eqnarray}
\langle k\rangle(s,t) = \left\{
\begin{array}{lcl}
\displaystyle\ln\left(\frac{t}{s}\right)+\mathcal{O}(1) & , & \alpha=0 \\
 & & \\
\displaystyle\left(1-\alpha\right)\ln\left(\frac{t}{s}\right)+\mathcal{O}(1) & , & 0<\alpha<1 \\
 & & \\
\displaystyle\ln\left(\frac{\ln t}{\ln s}\right)\left[1+o(1)\right] & , & \alpha=1 \\
 & & \\
2+\mathcal{O}(s^{-1}, \tau^{1-\alpha}) & , & \alpha>1
\end{array}
\right.\,.
\label{kmsmall}
\end{eqnarray}

On the other hand, if $1\ll s\ll t\ll m$ (in the sense also that $\ln\left(t/s\right)\gg 1$), one has
\begin{eqnarray}
\langle k\rangle(s,t) = \left\{
\begin{array}{lcl}
\displaystyle\ln\left(\frac{t}{s}\right)+\mathcal{O}(s^{-1}) & , & \alpha=0 \\
 & & \\
\displaystyle\frac{1-\alpha}{2^{1-\alpha}-1}\ln\left(\frac{t}{s}\right)+\mathcal{O}(1) & , & 0<\alpha<1 \\
 & & \\
\displaystyle\frac{1}{\ln 2}\ln\left(\frac{t}{s}\right)+\mathcal{O}(1) & , & \alpha=1 \\
 & & \\
\displaystyle\frac{\alpha-1}{1-2^{1-\alpha}}\ln\left(\frac{t}{s}\right)+\mathcal{O}(1) & , & 1<\alpha\leq 2 \\
 & & \\
\displaystyle\frac{\alpha-1}{1-2^{1-\alpha}}\ln\left(\frac{t}{s}\right)+\mathcal{O}(1) & , & \alpha>2 \quad\textrm{and}\quad m\gg t^{\alpha-1} \\
 & & \\
2 + \mathcal{O}(ms^{1-\alpha}) & , & \alpha>2 \quad\textrm{and}\quad m\ll s^{\alpha-1}
\end{array}
\right.
\label{kmbig}
\end{eqnarray}

The results for the mean degree in other regimes are presented in the Appendix \ref{md} for completeness. Note that the asymptotic behaviour of the mean degree for large number of branches is distinct from the one branch (or relatively small number of branches -- in the sense that $m\ll s$) case, showing that $m$ does influence on the graph property of the network. Basically, one has $\langle k\rangle(s,t)\sim\frac{1-\alpha}{2^{1-\alpha}-1}\ln\left(\frac{t}{s}\right)$ for any $\alpha$ for large $m$, except when the number of branches is large but not larger than $s^{\alpha-1}$ (see the $\alpha>2$ case). Actually, this last case shows that despite the fact of $m$ being large, it belongs to the case where the interaction parameter $\alpha$ is strong enough to let each vertex to have two links only: one from its ``own'' and the other from the vertex born immediately after it in the same branch; in other words, the interaction $\alpha$ overcomes the size of the network and each vertex can see its neighborhood only (therefore, when $\alpha$ is sufficiently large, the network effectively behaves as a small $m$ case).

As a last remark, the number of branches became important for non-zero $\alpha$ only, which leads the model to ``notice'' the metric structure of the space. In the particular case where $s\ll t$, in the sense that $\ln(t/s)\gg 1$, one has $\langle k\rangle(s,t)=\ln(t/s)+\mathcal{O}(1)$. These mean degrees recover results from recursive random graphs, where the metric structure that the system lies on can be ignored -- this is cleary seen from the fact that $w_{b}^{(m)}(s,t+1)=(mt+1)^{-1}$ for $\alpha=0$.


\section{Degree distribution}
\label{degree_distribution}

Consider first the case $m=1$. From the master equation (\ref{me1}), it is possible to show that (see Appendix \ref{dd} for technical details)
\begin{eqnarray}
\nonumber p(k,s,t) & = & \frac{1}{(k-1)!}\left[\sum_{r=s+1}^{t}w_{1}^{(1)}(s,r)\right]^{k-1}\exp\left[-\sum_{u=s+1}^{t}w_{1}^{(1)}(s,u)\right]\left[1+o(1)\right]
\label{pkst1}
\end{eqnarray}
if
\begin{eqnarray}
\left|w_{1}^{(1)}(s,u)\right| \ll 1\,, \quad \mbox{for }u\in(s,t]\subset\mathbb{Z}_{+}\,.
\label{condition}
\end{eqnarray}
This condition can be satisfied for $0\leq\alpha\leq 1$ if $s\gg 1$, but it does not hold for larger values of $\alpha$, except if $\alpha\sim 1^{+}$. The form of the equation for $p(k,s,t)$ given above shows a Poisson distribution with mean $\sum_{r=s+1}^{t}w_{1}^{(1)}(s,r)=\langle k\rangle(t)-1$ for $s<t$ (see (\ref{k1})). This means that the probability of a vertex $s$, at time $t$, having $k$ links is concentrated on the mean $\langle k\rangle(t)-1$, and the structure of the network is homogeneous.

Using the asymptotic form (\ref{pkst1}), one can show that the degree distribution $P(k):=\lim_{t\rightarrow\infty}P(k,t)$, with $P(k,t):=\sum_{s=1}^{t}p(k,s,t)/t$, behaves as
\begin{eqnarray}
P(k) = \left\{
\begin{array}{lcl}
\displaystyle\frac{1}{2-\alpha}\left(\frac{1-\alpha}{2-\alpha}\right)^{k-1} & , & 0\leq\alpha<1 \\
 & & \\
\displaystyle\frac{e^{-1}}{(k-1)!} & , & \alpha\sim1^{+}
\end{array}
\right.\,.
\label{Pk_m=1}
\end{eqnarray}
For $\alpha=1$, it is possible to evaluate the degree distribution at time $t$ as
\begin{eqnarray}
\nonumber P(k,t) & = & 2^{-k} \frac{\ln t}{t}\,_{k}F_{k}(2,\cdots,2;3,\cdots,3;\ln t) \\
                 & = & \frac{\ln t}{t}\sum_{n=0}^{\infty}\frac{1}{\left(n+2\right)^{k}}\frac{\left(\ln t\right)^{n}}{n!}\,, \quad \alpha=1\,,
\label{Pkt_a=1}
\end{eqnarray}
where $_{k}F_{k}(.,.;.)$ is the hypergeometric function.

A simple calculation shows that for $s\gg 1$, one has $\partial{p}/\partial\alpha<0$ for sufficiently large $k$. The degree distribution decays more rapidly as $\alpha$ increases. For instance, in the case $0\leq\alpha<1$, the decayment is exponential, and for $\alpha\sim 1^{+}$ the degree distribution decays more rapidly, as an inverse of a factorial. Moreover, in the limit of $\alpha\rightarrow\infty$, $P(k)$ tends to $\delta_{k,2}$.

For the other case, when the number of branches is large in the sense that $m\gg t$, a heuristic approach will be adopted to estimate the form of the degree distribution. If one assumes that the probability that a vertex $s$ has exactly $\langle k\rangle(s,t)$ links at time $t$ -- the ``$\delta$-Ansatz'' --, it is possible to show (using results for the mean degree $\langle k\rangle(s,t)$\cite{DM01, DM03}) that $P(k)$ has an exponential distribution
\begin{eqnarray}
P(k) \sim \frac{1}{K(\alpha)}e^{-k/K(\alpha)}\,,
\label{Pk}
\end{eqnarray}
whenever $\langle k\rangle(s,t)\sim K(\alpha)\ln(t/s)$, where $K(\alpha)$ is a factor that depends on $\alpha$. Assuming the ``$\delta$-Ansatz'', which turns to be reasonable in the regime $1\ll s\ll t$, the exponential form of the degree distribution suits the cases where $m\gg t\gg s\gg 1$ (for any non-negative $\alpha$), except when $\alpha$ is sufficiently large (in the sense that $ms^{1-\alpha}\ll 1$), when $P(k)$ tends to $\delta_{k,2}$.

When each vertex has just two links on average due to the high value of $\alpha$, the model indicates that each vertex has, apart from its ``own link'', another that comes from its successor in the same branch. Thus, the network achieves a regular structure, like a regular lattice, where many of the vertices have connection to its nearest neighbors only.

The metric nature of the graph is well characterized in the regime where the number of branches is not large like $m\gg t$. Despite the fact that the probability that decays algebrically with distance can tune the network between a random graph type one (low values of $\alpha$, typically $0\leq\alpha<1$) and a regular lattice type (large values of $\alpha$, when $\langle k\rangle(s,t)\sim 2$), if the number of branches is too large, the normalization (\ref{norm}) diverges with $m$ and $w_{b}^{(m)}(s,t)$ becomes ``nearly'' uniform. This is the origin of the observed randomness in the network connections, which reflects in the exponential form of the degree distribution (\ref{Pk}).


\section{Shortest path length}
\label{shortest_path_length}

This section will provide an estimation for the mean shortest path length $\ell(t)$ between two randomly chosen vertices, say, $x_{a}$ and $x_{b}$. Since they are randomly chosen, their distance from the central node will be taken as $\mathcal{O}(t)$.

In this work, the order of magnitude of the shortest path length between $x_{a}$ and $x_{b}$ will be considered to be the graph distance (path length) between $x_{a}$ and the central node $0$ plus the graph distance between $x_{b}$ and $0$. Denoting the (shortest) path between two vertices, $x_{1}$ and $x_{2}$ by $x_{1}\leftrightarrow x_{2}$, the estimation of the shortest path between $x_{a}$ and $x_{b}$ proposed in this work stems in replacing $x_{a}\leftrightarrow x_{b}$ by $x_{a}\leftrightarrow 0\leftrightarrow x_{b}$. This estimation is expected to be reasonable at least in the case where the number of branches is sufficiently large, because then the probability of the paths $x_{a}\leftrightarrow 0$ and $x_{b}\leftrightarrow 0$ having intersection other than the central node $0$ is low (meaning that the shortest path $x_{a}\leftrightarrow x_{b}$ includes, necesarilly, the vertex $0$). It will be shown later that if $m$ is sufficiently larger than $t$, the shortest path length is estimated as $\mathcal{O}(\ln t)$, which resembles the result from random graphs. Note that by construction, there are no two different shortest paths connecting two points in the model presented in this work, since a new born vertex links to only one old vertex.

\begin{figure}[htb]
\centering
\includegraphics[width=50.0mm, height=50.0mm]{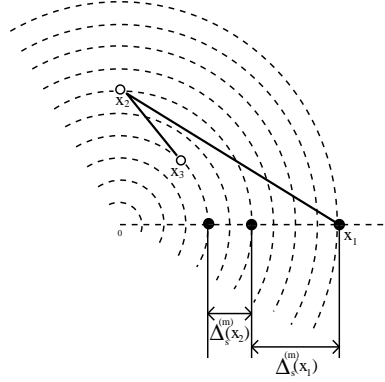}
\caption{Let the path $x_{1}\leftrightarrow x_{2}\leftrightarrow x_{3}\leftrightarrow\cdots$ be the shortest path between $x_{1}$ and the central node. The projection of $x_{2}$ and $x_{3}$ (and also $x_{1}$) onto the branch that joins $x_{1}$ and the central node are represented by a filled ball in the figure. The quantities $\Delta_{s}^{(m)}(x_{1})$ and $\Delta_{s}^{(m)}(x_{2})$ are also indicated. For sake of clearness, most of the branches were not drawn.}
\label{4}
\end{figure}

Next, define the function
\begin{eqnarray}
\Delta_{s}^{(m)}(r) := \sum_{s=1}^{r-1}\sum_{b=1}^{m}\left(r-s\right)w_{b}^{(m)}(s,r)=r-\sum_{s=1}^{r-1}\sum_{b=1}^{m}sw_{b}^{(m)}(s,r)\,,
\label{Deltas}
\end{eqnarray}
which is the measure of the mean (metric) distance covered by a link (starting from a point at distance $r$ from the origin) \textit{toward} the central node. In other words, this is the metric length of the projection of a link on some fixed branch (see Figure \ref{4}).

The evaluation of $\Delta_{s}^{(m)}(r)$ will focus on two distinct limit situations for the number of branches, and it will play a major role in the estimation of the shortest path length. As mentioned earlier, one should estimate the (order of the) number of paths that a randomly chosen vertex $x_{a}$ takes to achieve the central node (and the same for $x_{b}$); furthermore, as stated in the beginning of this section, it will be assumed that $x_{a}=\mathcal{O}(t)$ (the distance between $x_{a}$ and $0$) and by the same argument, one has also $x_{b}=\mathcal{O}(t)$. This means that the path length $x_{a}\leftrightarrow 0\leftrightarrow x_{b}$ (which is an estimation of the shortest path length $x_{a}\leftrightarrow x_{b}$) is of the order of the path length $x_{a}\leftrightarrow 0$ in this work.

Let $x_{a}=x_{1}\leftrightarrow x_{2}\leftrightarrow\cdots\leftrightarrow x_{n}=0$ be the shortest path between $x_{a}=x_{1}$ and the central node $0=x_{n}$. Using (\ref{Deltas}), one has
\begin{eqnarray}
\sum_{i=1}^{n-1}\Delta_{s}^{(m)}(x_{i}) = \textrm{distance between $x_{a}$ and $0$, which is $\mathcal{O}(t)$}\,.
\label{sum_Delta}
\end{eqnarray}
Remembering that $n$ is of the order of the shortest path length $\ell(t)$ between $x_{a}$ and $x_{b}$ at time $t$, the relation (\ref{sum_Delta}) can be used to estimate $\ell(t)$.

The details of the estimation is presented in Appendix \ref{app_spl}. For $m\ll s$, when the number of branches is not large, one has

\begin{center}
\begin{tabular}{|c|c|}
\hline
$\alpha$           & $\ell(t)$                     \\ \hline\hline
$0\leq\alpha\leq1$ & $\mathcal{O}(\ln t)$          \\ \hline
$1<\alpha<2$       & $\mathcal{O}(t^{\alpha-1})$   \\ \hline
$\alpha=2$         & $\mathcal{O}(t/\ln t)$        \\ \hline
$\alpha>2$         & $\mathcal{O}(t)$              \\ \hline
\end{tabular}
\end{center}
On the other hand, if $m\gg t$, one has
\begin{center}
\begin{tabular}{|c|c|c|}
\hline
$\alpha$           & $m$, $t$                              & $\ell(t)$                     \\ \hline\hline
$0\leq\alpha\leq2$ &  ---                                  & $\mathcal{O}(\ln t)$          \\ \hline
$\alpha>2$,        & $mt^{1-\alpha}\gg 1$                   & $\mathcal{O}(\ln t)$          \\ \hline
$2<\alpha<3$,      & $mt^{1-\alpha}\ll 1$                   & $\mathcal{O}(t^{\alpha-2}/m)$  \\ \hline
$\alpha=3$,        & $mt^{1-\alpha}\ll 1$                   & $\mathcal{O}(t^{2}/(m\ln t))$ \\ \hline
$\alpha>3$,        & $mt^{1-\alpha}\ll 1\ll mt^{2-\alpha}$   & $\mathcal{O}(t^{\alpha-2}/m)$  \\ \hline
$\alpha>3$,        & $mt^{2-\alpha}\ll 1$                   & $\mathcal{O}(t)$              \\ \hline
\end{tabular}
\end{center}

The above results indicate that whenever the number of branches is too large (in the sense of $mt^{1-\alpha}\gg 1$), the model behaves as a random graph, leading to a shortest path length between two randomly chosen vertex that increases logarithmically with $t$.


\section{Conclusions}
\label{conclusions}

The present work has introduced a network growing toward $m$ branches embedded on a metric structure and analysed some graph properties. The mean degree of a fixed vertex decreases as $\alpha$ increases, indicating, as expected, that the strength of the ``interaction'' confines the vertices to link to other vertices that are located nearer than farther. To ensure the fact that the number $m$ of branches does influence on the graph properties like mean degree, two opposite conditions were then considered. For small $m$, the results were similar to the one branch case, while for large $m$, the mean degree was larger. In any case, for $\alpha$ sufficiently large, one has $\langle k\rangle(s,t)\rightarrow 2$, no matter the number of branches. The degree distribution was evaluated for $0\leq\alpha<1$ and $\alpha\sim 1^{+}$, and it shows, as expected, that they decay as rapidly as the magnitude of $\alpha$ increases. Finally, the shortest path length between two randomly chosen vertices was estimated as a function of $t$. The metric structure of the model is better displayed if the number of branches are not overwhelmingy large; otherwise, the model behaves as a random graph, and this fact is supported by the results obtained for the degree distribution, shortest path length and mean degree; this last one grows as $\ln(t/s)$, when $s\ll t$, for any $\alpha$ for sufficiently large $m$ -- see (\ref{kmbig}) --, as it is expected\cite{DM03}. The model presented in this work was successful in displaying properties that are common to networks that are embedded in a Euclidean space (like $\mathbb{R}^{2}$) and is manageable to analytic treatment.


\section{Acknowledgements}
\label{acknowledgements}

The authors are thankful to S. N. Dorogovtsev for comments and MOH is supported by the project DYSONET.


\renewcommand{\thesection}{\Alph{section}}
\setcounter{section}{0}

\section{Appendix}

This appendix is devoted to present some technical observations and list results omitted in the main text for clearness.

The present work has used extensively the fact that
\begin{eqnarray}
\sum_{r=1}^{t-1}r^{-\alpha} \sim \left\{
\begin{array}{lcl}
\displaystyle\frac{t^{1-\alpha}}{1-\alpha} + \zeta(\alpha) + \mathcal{O}(t^{-\alpha}) & , & 0 \leq\alpha < 1 \\
 & & \\
\ln t + \gamma + \mathcal{O}(t^{-1})                                      & , & \alpha = 1 \\
 & & \\
\zeta(\alpha) - \displaystyle\frac{t^{1-\alpha}}{\alpha-1} + \mathcal{O}(t^{-\alpha}) & , & \alpha > 1
\end{array}
\right.\,,
\label{asymp}
\end{eqnarray}
where $\zeta$ is the zeta function of Riemann and $\gamma$ ($=0.5772...$) is the Euler - Mascheroni constant. As a remark, $\zeta(\alpha)>0$ for $\alpha>1$ and $\zeta(\alpha)<0$ for $\alpha<1$.

\subsection{One branch case}
\label{onebranchcase}

For $1\ll s<t$, one has
\begin{eqnarray}
\langle k\rangle(s,t) = \left\{
\begin{array}{l}
1 + \displaystyle\left(1-\alpha\right)\int\limits_{s/t}^{1}\frac{dr}{r\left(1-r\right)^{\alpha}} + \mathcal{O}(s^{\alpha-1})\,, \\
\hfill 0\leq\alpha<1 \\
1 + \displaystyle\int\limits_{s+1}^{t}\frac{dr}{\left(r-s\right)\ln r} + \mathcal{O}(\ln^{-1}s)\,, \\
\hfill \alpha=1 \\
1 + \displaystyle\frac{1}{\zeta(\alpha)}\sum_{u=1}^{\tau}u^{-\alpha} + \mathcal{O}(s^{1-\alpha})\,, \\
\hfill \alpha>1
\end{array}
\right.\,,
\label{1_ll_st}
\end{eqnarray}
where $\tau := t-s$. 

The case $\alpha=1$ may deserve a short comment. Consider the case $\alpha=1$, $m=1$ and $1\ll s\ll t$ such that $t-s\gg s^{\delta}$ for some positive $\delta=\mathcal{O}(1)$. Using (\ref{asymp}), the mean degree (\ref{k1}) is
\begin{eqnarray}
\nonumber \langle k\rangle(s,t) & = & 1 + \sum_{r=s+1}^{t}\frac{\left(r-s\right)^{-1}}{\ln r + \mathcal{O}(1)} \\
\nonumber & = & \int\limits_{1}^{t-s} \frac{dy}{y\ln\left(y+s\right)} + \mathcal{O}\left(1, \int\limits_{1}^{t-s} \frac{dy}{y\ln^{2}\left(y+s\right)}\right)\,. \\
\label{a1}
\end{eqnarray}

However, since
\begin{eqnarray}
\nonumber \left|\int\limits_{1}^{t-s} \frac{dy}{y\ln^{2}\left(y+s\right)}\right| & \leq & \int\limits_{1}^{e} \frac{dz}{z\ln^{2}s} + \int\limits_{e}^{t-s} \frac{dz}{z\ln^{2}z} \\
\nonumber & = & \mathcal{O}(\ln^{-2}s) + \int\limits_{1}^{\ln(t-s)}\frac{du}{u^{2}} = \mathcal{O}(1)\,, \\
\label{a2}
\end{eqnarray}
one is left with
\begin{eqnarray}
\nonumber \langle k\rangle(s,t) & = & \int\limits_{1}^{t-s}\frac{dy}{y\ln\left(y+s\right)} + \mathcal{O}(1) \\
\nonumber & = & \int\limits_{1}^{s^{\delta+1}} \frac{dy}{y\ln\left(y+s\right)} + \int\limits_{s^{\delta+1}}^{t-s} \frac{dy}{y\ln\left(y+s\right)} + \mathcal{O}(1)\,, \\
\label{a3}
\end{eqnarray}
where the first term (last line) on the right hand is bounded ($\delta=\mathcal{O}(1)$):
\begin{eqnarray}
\left|\int\limits_{1}^{s^{\delta+1}} \frac{dy}{y\ln\left(y+s\right)}\right| & \leq & \frac{1}{\ln s} \int\limits_{1}^{s^{\delta+1}} \frac{dy}{y} = \mathcal{O}(1)\,.
\label{a4}
\end{eqnarray}
Thus,
\begin{eqnarray}
\nonumber \langle k\rangle(s,t) & = & \int\limits_{s^{\delta+1}}^{t-s} \frac{dy}{y\ln\left(y+s\right)} + \mathcal{O}(1) \\
\nonumber & = & \int\limits_{1+s^{\delta}}^{t/s} \frac{dz}{z\ln(sz)}\left(1+\frac{1}{z-1}\right) + \mathcal{O}(1) \\
 & = & \int\limits_{1+s^{\delta}}^{t/s} \frac{dz}{z\ln(sz)}\Big[1+\mathcal{O}(s^{-\delta})\Big] + \mathcal{O}(1)\,,
\label{a5}
\end{eqnarray}
and the desired result follows.

The estimation for $\alpha=1$ assumed that $\tau\gg s^{\delta}$ for some positive $\delta=\mathcal{O}(1)$.


\subsection{On the normalization (\ref{norm})}
\label{app_normalization}

The (asymptotic form of) normalization factor is evaluated by considering the cases $m$ odd and $m$ even separately. It is shown that both have the same general formula.

For $m$ odd, the normalization is
\begin{eqnarray}
\nonumber N_{odd}(r,m,\alpha) & = & r^{-\alpha} + \sum_{s=1}^{r-1}\left(r-s\right)^{-\alpha} + 2\sum_{s=1}^{r-1}\sum_{j=1}^{\frac{m-1}{2}}\Big[r-s+\min\{2s,j\}\Big]^{-\alpha}\,, \\
\label{a10}
\end{eqnarray}
where the first term is the distance between a $r$ born vertex to the central node; the second term counts the distance between this new vertex to the vertices located in the same branch, and the last term the remaining ones.

If $m$ is even, one has
\begin{eqnarray}
\fl\nonumber N_{even}(r,m,\alpha) = r^{-\alpha} + \sum_{s=1}^{r-1}\left(r-s\right)^{-\alpha} + \sum_{s=1}^{r-1}\Big[r-s+\min\{2s,m/2\}\Big]^{-\alpha} + \\
 + 2\sum_{s=1}^{r-1}\sum_{j=1}^{\frac{m}{2}-1}\Big[r-s+\min\{2s,j\}\Big]^{-\alpha}\,,
\label{a11}
\end{eqnarray}
where the meaning of each term is similar to the odd case. The third term is the contribution due to the branch that diametrically opposes branch $1$, which is absent if $m$ is odd.

For $1<m\ll r$, one has
\begin{eqnarray}
N_{\left\{{odd \atop even}\right\}}(r,m,\alpha) = \left\{
\begin{array}{lcl}
\displaystyle\frac{mr^{1-\alpha}}{1-\alpha} + C(m,\alpha) + \mathcal{O}(r^{-\alpha}) & , & 0\leq\alpha<1 \\
 & & \\
m\ln r + C(m,1) + \mathcal{O}(r^{-1}) & , & \alpha=1 \\
 & & \\
C(m,\alpha) - \displaystyle\frac{mr^{1-\alpha}}{\alpha-1} + \mathcal{O}(r^{-\alpha}) & , & \alpha>1
\end{array}
\right.\,,
\label{a12}
\end{eqnarray}
where $C(m,\alpha)$ is defined in (\ref{c}).

For $m\gg r$, one has
\begin{eqnarray}
\fl \nonumber N_{\left\{{odd \atop even}\right\}}(r,m,\alpha) = \\
\fl = \left\{
\begin{array}{lcl}
\displaystyle\left(\frac{2^{1-\alpha}-1}{1-\alpha}\right)mr^{1-\alpha} - \left(\frac{1+2^{-\alpha}}{2}\right)mr^{-\alpha} + \frac{8\left[\left(1+\alpha\right)2^{-\alpha}-1\right]}{\left(1-\alpha\right)\left(2-\alpha\right)}r^{2-\alpha} +  & & \\
 & & \\
 + \mathcal{O}(mr^{-\alpha-1}, r^{1-\alpha}) & , & 0\leq\alpha<1 \\
 & & \\
m\ln 2 - \displaystyle\frac{3}{4}mr^{-1} + 4\left(2\ln 2-1\right)r + \mathcal{O}(\ln r, mr^{-2}) & , & \alpha=1 \\
 & & \\
\displaystyle\left(\frac{1-2^{1-\alpha}}{\alpha-1}\right)mr^{1-\alpha} - \left(\frac{1+2^{-\alpha}}{2}\right)mr^{-\alpha} + \frac{8\left[1-\left(1+\alpha\right)2^{-\alpha}\right]}{\left(\alpha-1\right)\left(2-\alpha\right)}r^{2-\alpha} + & & \\
 & & \\
 + \mathcal{O}(mr^{-\alpha-1}, 1) & , & 1<\alpha<2 \\
 & & \\
\displaystyle\frac{1}{2}mr^{-1} - \frac{5}{8}mr^{-2} + 2\ln r + 4 + 2\gamma - \zeta(2) - 6\ln 2 + \mathcal{O}(mr^{-3}, r^{-1}) & , & \alpha=2 \\
 & & \\
\displaystyle\left(\frac{1-2^{1-\alpha}}{\alpha-1}\right)mr^{1-\alpha} - \left(\frac{1+2^{-\alpha}}{2}\right)mr^{-\alpha} + 2\zeta(\alpha-1) - \zeta(\alpha) + & & \\
 & & \\
 + \displaystyle\frac{8\left[\left(1+\alpha\right)2^{-\alpha}-1\right]}{\left(\alpha-1\right)\left(\alpha-2\right)}r^{2-\alpha} + \mathcal{O}(mr^{-\alpha-1}, r^{1-\alpha}) & , & \alpha>2
\end{array}
\right.\,.
\label{a13}
\end{eqnarray}


\subsection{On the $m$ branch master equation}
\label{mmeandegree}

After multiplying both sides of the master equation (\ref{me}) by $k$ and summing them, one gets
\begin{eqnarray}
\langle k\rangle(s,t+1) = A\langle k\rangle(s,t) + B(s,t+1)\,.
\label{a6}
\end{eqnarray}
Since one can show that
\begin{eqnarray}
\fl\nonumber A := \displaystyle\prod_{j=1}^{m}w_{j}^{(m)}(s,t+1) + \prod_{j=1}^{m}\overline{w}_{j}^{(m)}(s,t+1) + \\
\nonumber + \sum_{b=1}^{m-1}\sum_{q_{1}<\cdots<q_{b}} w_{q_{1}}^{(m)}(s,t+1) \cdots w_{q_{b}}^{(m)}(s,t+1) \prod_{\stackrel{j=1}{j\neq q_{1},\cdots,q_{b}}}^{m}\overline{w}_{j}^{(m)}(s,t+1) \\
 = 1
\label{a7}
\end{eqnarray}
and
\begin{eqnarray}
\fl\nonumber B(s,t+1) := m\displaystyle\prod_{j=1}^{m}w_{j}^{(m)}(s,t+1) + \sum_{b=1}^{m-1} b \sum_{q_{1}<\cdots<q_{b}}w_{q_{1}}^{(m)}(s,t+1)\cdots w_{q_{b}}^{(m)}(s,t+1) \times \\
\nonumber \times\prod_{\stackrel{j=1}{j\neq q_{1},\cdots,q_{b}}}^{m} \overline{w}_{j}^{(m)}(s,t+1) \\
 = \sum_{j=1}^{m} w_{j}^{(m)}(s,t+1)
\label{a8}
\end{eqnarray}
for all positive integer $m$, the results follows. The formula (\ref{a7}) can be easily obtained if one notices that $A = \prod_{j=1}^{m}\Big[w_{j}^{(m)}(s,t+1)+\overline{w}_{j}^{(m)}(s,t+1)\Big]$. On the other hand, although the deduction of the formula (\ref{a8}) is straighforward, it is tedious. It may be of some help to introduce the quantity $r_{q}^{(m)}(s,t) := w_{q}^{(m)}(s,t)/\overline{w}_{q}^{(m)}(s,t)$, which is well defined (since $\overline{w}_{q}(s,t)$ does not go to zero), and deduce the formula by induction by working out the expression
\begin{eqnarray}
\fl B(s,t+1) = m\prod_{j=1}^{m}w_{j}^{(M)}(s,t+1) + \left[\prod_{j=1}^{m} \overline{w}_{j}^{(M)}(s,t+1)\right] \times \\
\times\sum_{b=1}^{m-1} b \sum_{q_{1}<\cdots<q_{b}} r_{q_{1}}^{(M)}(s,t+1)\cdots r_{q_{b}}^{(M)}(s,t+1)\,.
\label{a9}
\end{eqnarray}
Note that the superscript of the probability $w_{j}^{(m)}(s,t+1)$ was replaced by a fixed number $M$. The induction argument will be applied to the variable $m$ only, and it can be shown that it is valid for any $M$ -- in particular, in the case of interest $M=m$. The reason for make $M$ independent of $m$ is twofold. First, the expression (\ref{a8}) is valid for any $w_{j}^{(m)}(s,t+t)$ and $\overline{w}_{j}^{(m)}(s,t+1)$ that satisfies $w_{j}^{(m)}(s,t+1)+\overline{w}_{j}^{(m)}(s,t+1)=1$ for each $j$, and does not depend on the particular choice of $w_{j}^{(m)}(s,t+1)$ and $\overline{w}_{j}^{(m)}(s,t+1)$ like (\ref{w}); to define $r_{q}^{(M)}(s,t+1)$, one should impose the additional condition $\overline{w}_{j}^{(M)}(s,t+1)\neq 0$. Secondly, without making $M$ independent of $m$ (in other words, if one keeps $m$ instead of $M$), the induction argument became extremely tedious to be applied, since the function $w_{j}^{(m)}(s,t+1)$, from the induction hypothesis, is different from the $w_{j}^{(m+1)}(s,t+1)$ (see the definitions (\ref{w}) and (\ref{dstb})).


\subsection{Mean degree}
\label{md}

Before analysing the mean degree for many cases not presented in the main text, it is convenient to define
\begin{eqnarray}
C(m,\alpha):=\left\{
\begin{array}{l}
\left.
\begin{array}{lcl}
m\zeta(\alpha)-2\displaystyle\sum_{b=1}^{\frac{m-1}{2}}\sum_{u=1}^{b}u^{-\alpha} & , & m\textrm{ odd} \\
 & & \\
m\zeta(\alpha)-\displaystyle\sum_{u=1}^{m/2}u^{-\alpha} - 2\displaystyle\sum_{b=1}^{\frac{m}{2}-1}\sum_{u=1}^{b}u^{-\alpha} & , & m\textrm{ even}
\end{array}
\right\} \alpha\neq 1
 \\ \\
\left.
\begin{array}{lcl}
m\gamma-2\displaystyle\sum_{b=1}^{\frac{m-1}{2}}\sum_{u=1}^{b}u^{-1} & , & m\textrm{ odd} \\
 & & \\
m\gamma-\displaystyle\sum_{u=1}^{m/2}u^{-1} - 2\displaystyle\sum_{b=1}^{\frac{m}{2}-1}\sum_{u=1}^{b}u^{-1} & , & m\textrm{ even}
\end{array}
\right\} \alpha=1
\end{array}
\right.\,\hspace{-5.8mm}.
\label{c}
\end{eqnarray}

This work has concentrated in two limiting regimes to stress the role played by $m$. The results for ``small'' number of branches ($1<m\ll s<t$) was compared to the other limit case, where $m$ is ``large'' ($1\ll s<t\ll m$). In both cases, $s$ is chosen to be much larger than $1$ and in this appendix, $t(>s)$ is taken free, although simple analytic results are provided if $s\ll t$, as shown in the section \ref{m_branches}.

For $1<m\ll s<t$, one has
\begin{eqnarray}
\fl
\nonumber \langle k\rangle(s,t)=\\
\fl = \left\{
\begin{array}{lcl}
1+\displaystyle\ln\left(\frac{t}{s}\right)+\mathcal{O}(s^{-1})\,, \quad \alpha=0 \\
 \\
1+\left(1-\alpha\right)\displaystyle\int\limits_{s/t}^{1}\frac{dy}{y\left(1-y\right)^{\alpha}} + \mathcal{O}(s^{\alpha-1})\,, \quad 0<\alpha<1 \\
 \\
1+\displaystyle\int\limits_{s+1}^{t}\frac{dy}{\left(y-s\right)\ln y} + \mathcal{O}(\ln^{-1}s)\,, \quad \alpha=1 \\
 \\
1+\displaystyle\frac{m\zeta(\alpha)}{C(m,\alpha)}-\frac{1}{C(m,\alpha)}\Bigg[ 2\sum_{b=1}^{\frac{m-1}{2}}\sum_{u=1}^{b}u^{-\alpha} + \sum_{u=\tau+1}^{\infty}u^{-\alpha} + 2\sum_{b=1}^{\frac{m-1}{2}}\sum_{u=\tau+b+1}^{\infty}u^{-\alpha} \Bigg] + \\
 \\
 + \mathcal{O}(s^{1-\alpha}) ,\quad m\textrm{ odd}\,, \quad \alpha>1 \\
 \\
1+\displaystyle\frac{m\zeta(\alpha)}{C(m,\alpha)}-\frac{1}{C(m,\alpha)}\Bigg[ 2\sum_{b=1}^{\frac{m}{2}-1}\sum_{u=1}^{b}u^{-\alpha} + \sum_{u=1}^{m/2}u^{-\alpha} + 2\sum_{b=1}^{\frac{m}{2}-1}\sum_{u=\tau+b+1}^{\infty}u^{-\alpha} + \\
 \\
 + \displaystyle\sum_{u=\tau+1}^{\infty}u^{-\alpha} + \sum_{u=\tau+m/2+1}^{\infty}u^{-\alpha} \Bigg] + \mathcal{O}(s^{1-\alpha}) ,\quad m\textrm{ even}\,, \quad \alpha>1
\end{array}
\right.\,,
\label{1<m<<s<t}
\end{eqnarray}
where $\tau:=t-s$. The calculations can not be always computed directly in the $\alpha>1$ case, and the mean degree is estimated by evaluating bounds. However, in many cases the lower and upper bounds have the same order of magnitude with the same leading term. The results in (\ref{1<m<<s<t}) show that the behaviour of the mean degree for $1<m\ll s<t$ is similar to the one branch case. 

On the other hand, for $1\ll s<t\ll m$, one has
\begin{eqnarray}
\fl
\nonumber \langle k\rangle(s,t)=\\
\fl = \left\{
\begin{array}{lcl}
1+\displaystyle\ln\left(\frac{t}{s}\right)+\mathcal{O}(s^{-1}) & , & \alpha=0 \\
 \\
1 + \displaystyle\frac{1-\alpha}{2^{1-\alpha}-1}\int\limits_{s/t}^{1}\frac{dy}{y\left(1+y\right)^{\alpha}} + \mathcal{O}\big(s^{-1}, t/m\big) & , & 0<\alpha<1 \\
 \\
1 + \displaystyle\frac{1}{\ln 2}\ln\left(\frac{t+s}{2s}\right) + \mathcal{O}\Big( s^{-1}, t/m, (s/m)\ln t, (\tau/m)\ln s \Big) & , & \alpha=1 \\
1+\displaystyle\frac{\alpha-1}{1-2^{1-\alpha}}\int\limits_{s/t}^{1}\frac{dy}{y\left(1+y\right)^{\alpha}} + \left\{
\begin{array}{lcl}
\mathcal{O}\displaystyle\left(\frac{t}{m}, \frac{1}{s}\right) & , & 1<\alpha<2 \\
 & & \\
\mathcal{O}\displaystyle\left(\frac{t\ln t}{m}, \frac{1}{s}\right) & , & \alpha=2 \\
 & & \\
\mathcal{O}\left(\displaystyle\frac{t^{\alpha-1}}{m}, \frac{1}{s}\right)\,, & & \alpha>2
\end{array}
\right.
\end{array}
\right.\,\hspace{-2.3mm},
\label{1<<s<t<<m}
\end{eqnarray}
where it was assumed that $s\ln t\ll m$ and $\tau\ln t\ll m$ for $\alpha=1$; $t\ln t\ll m$ for $\alpha=2$; and $t^{\alpha-1}\ll m$ for $\alpha>2$.

For $1\ll s<t\ll m$, the interplay between the free parameters should be examined, specially the competition between the strength of ``interaction'' $\alpha$ and the number of branches $m$. The present work has considered the limiting cases $t^{\alpha-1} \ll m$ and $s^{\alpha-1} \gg m$ to show their differences. Since $m\gg t$, note that the former condition (``large $m$'' -- in contrast to the later, $s^{\alpha-1}\gg m$, the ``large $\alpha$'') is always satisfied for $0\leq\alpha\leq 2$. Finally, the parameters $mt^{1-\alpha}$ and $ms^{\alpha-1}$ dictates the behaviour of the mean degree in the case $\alpha>2$. If $t^{\alpha-1} \ll m$, one has the situation shown in (\ref{1<<s<t<<m}), where for sufficiently large $m$ (such that $m\gg t^{\alpha-1}$), the mean degree grows with the size of network. On the other hand, if $s^{\alpha-1}\gg m$, or, equivalently, the $\alpha$ is sufficiently strong, one has (\ref{1<m<<s<t}), where the mean degree behaves as in the case $m=1$.


\subsection{On the degree distribution ($m=1$)}
\label{dd}

Performing the zeta transform
\begin{eqnarray}
\tilde{p}(K,s,t) := \sum_{k=1}^{\infty}K^{k}p(k,s,t)
\label{a15}
\end{eqnarray}
on the master equation (\ref{me1}), one can show that
\begin{eqnarray}
\nonumber \tilde{p}(K,s,t+1) & = & \tilde{p}(K,s,s)\prod_{u=s}^{t}\frac{\tilde{p}(K,s,u+1)}{\tilde{p}(K,s,u)} \\
\nonumber & = & K\prod_{u=s}^{t}\left[1 + \left(K-1\right)w(s,u+1)\right]\,, \\
\label{a16}
\end{eqnarray}
since $\tilde{p}(K,s,u)\neq 0$ for any $u\geq s$. From the condition $p(k,s,s)=\delta_{k,1}$ (or $\tilde{p}(K,s,s)=K$), one has
\begin{eqnarray}
\fl\nonumber p(k,s,t+1) = \frac{1}{(k-1)!}\sum_{q=0}^{k-1}\left(k-1 \atop q\right)\times \\
\times \left\{\frac{d^{k-1-q}}{dK^{k-1-q}}\left.\exp\left[\left(K-1\right)\sum_{u=s}^{t}w(s,u+1)\right]\right|_{K=0}\right\} \left\{T^{q}(\alpha,K)\Bigg|_{K=0}\right\}\,,
\label{a17}
\end{eqnarray}
where
\begin{eqnarray}
T^{q}(\alpha,K)\Bigg|_{K=0} \hspace{-1.79mm} = \delta_{q,0} + \left\{
\begin{array}{lcl}
\mathcal{O}\left(q!s^{2(\alpha-1)}\right) & , & 0\leq\alpha<1 \\
 & & \\
\mathcal{O}\left(\displaystyle\frac{q!}{\ln^{2}s}\right) & , & \left\{
\begin{array}{l}
\alpha = 1 \\
 \\
\alpha \rightarrow 1^{+}
\end{array}
\right. \\
\end{array}
\right.
\label{a18}
\end{eqnarray}
for sufficielntly large $s$. Therefore, for large $s$ (say, $s\gg e^{\sqrt{k!}}$), one has
\begin{eqnarray}
\nonumber p(k,s,t) & = &\frac{e^{-\left[\langle k\rangle(s,t)-1\right]}}{(k-1)!}\left[\langle k\rangle(s,t)-1\right]^{k-1} \left[1+o(1)\right]\,, \\
\label{a19}
\end{eqnarray}
where the main contribution comes from the $q=0$ term (in the sum (\ref{a17})) and $\langle k\rangle(s,t)$ is given by (\ref{k1}).


\subsection{Shortest path length}
\label{app_spl}

Firstly, the asymptotic form of $\Delta_{s}^{(m)}(r)$, which is closely related to the estimation of the shortest path length, will be listed. The definition (\ref{c}) will be used below.

For $1<m\ll r$, one has
\begin{eqnarray}
\Delta_{s}^{(m)}(r) = \left\{
\begin{array}{lcl}
\displaystyle\left(\frac{1-\alpha}{2-\alpha}\right)r + \mathcal{O}(r^{\alpha}) & , & 0\leq\alpha<1 \\
 & & \\
\displaystyle\frac{r}{\ln r} + \mathcal{O}\left(\frac{r}{\ln^{2}r}\right) & , & \alpha=1 \\
 & & \\
\displaystyle\frac{mr^{2-\alpha}}{C(m,\alpha)\left(2-\alpha\right)} + \mathcal{O}(r^{3-2\alpha}, 1) & , & 1<\alpha<2 \\
 & & \\
\displaystyle\frac{m}{C(m,2)}\ln r + \mathcal{O}(1) & , & \alpha=2 \\
\end{array}
\right.
\label{Deltas1<m_ll_r}
\end{eqnarray}
and
\begin{eqnarray}
\fl \nonumber \Delta_{s}^{(m)}(r) = \frac{C(m,\alpha-1)}{C(m,\alpha)} + \\
\fl + \left\{
\begin{array}{lcl}
\displaystyle\frac{1}{C(m,\alpha)}\sum_{u=1}^{\frac{m-1}{2}}\left(\frac{m^{2}-1}{4}+u-u^{2}\right)u^{-\alpha} - & & \\
 - \frac{\zeta(\alpha)}{C(m,\alpha)}\left(\frac{m^{2}-1}{4}\right) & , & m \textrm{ odd} \\
 & & \\
\displaystyle\frac{1}{C(m,\alpha)}\sum_{u=1}^{\frac{m}{2}-1}\left(\frac{m^{2}-2m}{4}+u-u^{2}\right)u^{-\alpha} + \\
 & & \\
 + \displaystyle\frac{m}{2}\left[\sum_{u=1}^{m/2}u^{-\alpha} - \zeta(\alpha)\right] - \frac{\zeta(\alpha)}{C(m,\alpha)}\frac{m}{2}\left(\frac{m}{2}-1\right) & , & m \textrm{ even}
\end{array}
\right\} + \mathcal{O}(r^{2-\alpha})
\label{Deltas1<m_ll_r_a>2}
\end{eqnarray}
for $\alpha>2$.
Note that $\Delta_{s}^{(m)}(r)$ becomes smaller as it approaches the central node.

On the other hand, for $1\ll r\ll m$, the parameter $mr^{1-\alpha}$ and $mr^{2-\alpha}$ play a major rule to the asymptotic behaviour of $\Delta_{s}^{(m)}(r)$. Firstly, one can show that
\begin{eqnarray}
\Delta_{s}^{(m)}(r) = \left\{
\begin{array}{lcl}
\displaystyle\frac{\left(\alpha+2^{2-\alpha}-3\right)r}{\left(2-\alpha\right)\left(2^{1-\alpha}-1\right)} + \mathcal{O}\left(1, \frac{r^{2}}{m}\right) & , & 0\leq\alpha<1 \\
 & & \\
\displaystyle\left(2 - \frac{1}{\ln 2}\right)r + \mathcal{O}\left(1, \frac{r^{2}}{m}\right) & , & \alpha=1 \\
 & & \\
\displaystyle\frac{\left(\alpha+2^{2-\alpha}-3\right)r}{\left(2-\alpha\right)\left(2^{1-\alpha}-1\right)} + \mathcal{O}\left(1, \frac{r^{2}}{m}\right) & , & 1<\alpha<2 \\
 & & \\
\displaystyle\left(2-2\ln 2\right)r + \mathcal{O}\left(1, \frac{r^{2}\ln r}{m}\right) & , & \alpha=2
\end{array}
\right.\,,\\
\label{Deltas1_ll_r_ll_m}
\end{eqnarray}
where a stronger condition, $m\gg r\ln r$, was assumed for $\alpha=2$. For $\alpha>2$ and $mr^{1-\alpha}\gg 1$, one has
\begin{eqnarray}
\Delta_{s}^{(m)}(r) = \left\{
\begin{array}{lcl}
\displaystyle\frac{\left(\alpha+2^{2-\alpha}-3\right)r}{\left(2-\alpha\right)\left(2^{1-\alpha}-1\right)} + \mathcal{O}\left(1, \frac{r^{\alpha}}{m}\right) & , & 2\leq\alpha<3 \\
 & & \\
\displaystyle\frac{2}{3}r + \mathcal{O}\left(1, \frac{r^{3}}{m}\right) & , & \alpha=3 \\
 & & \\
\displaystyle\frac{\left(\alpha+2^{2-\alpha}-3\right)r}{\left(2-\alpha\right)\left(2^{1-\alpha}-1\right)} + \mathcal{O}\left(1, \frac{r^{\alpha}}{m}\right) & , & \alpha>3
\end{array}
\right.\,.\\
\label{regime_i}
\end{eqnarray}
In this regime, one sees that the leading term of $\Delta_{s}^{(m)}(r)$ (for large $r$) is proportional to $r$. Note that the condition $mr^{1-\alpha}\gg 1$ is automatically satisfied for $0\leq\alpha\leq 2$ (with $m\gg r$). A different asymptotic behaviour is seen for $mr^{1-\alpha}\ll 1$; if $2<\alpha\leq 3$, one gets
\begin{eqnarray}
\fl\Delta_{s}^{(m)}(r) = \left\{
\begin{array}{lcl}
\displaystyle\frac{\left(\alpha+2^{2-\alpha}-3\right)}{\left(2-\alpha\right)\left(1-\alpha\right)}\frac{mr^{2-\alpha}}{\left[2\zeta(\alpha-1)-\zeta(\alpha)\right]} + & & \\
 & & \\
 + \mathcal{O}\left(m^{2}r^{3-2\alpha}, mr^{1-\alpha}, r^{3-\alpha}\right) & , & 2<\alpha<3 \\
 & & \\
\displaystyle\frac{1}{4}\frac{mr^{-1}}{\left[2\zeta(2)-\zeta(3)\right]} + \mathcal{O}\left(m^{2}r^{-3}, mr^{-1}, \ln r\right)& , & \alpha=3
\end{array}
\right.\,,
\label{2<a<3}
\end{eqnarray}
and for $\alpha>3$, one has
\begin{eqnarray}
\fl\Delta_{s}^{(m)}(r) = \left\{
\begin{array}{lcl}
\displaystyle\frac{\left(\alpha+2^{2-\alpha}-3\right)}{\left(2-\alpha\right)\left(1-\alpha\right)}\frac{mr^{2-\alpha}}{\left[2\zeta(\alpha-1)-\zeta(\alpha)\right]} + & & \\
 & & \\
 + \mathcal{O}\left(m^{2}r^{3-2\alpha}, 1\right) & , & mr^{1-\alpha}\ll 1\ll mr^{2-\alpha} \\
 & & \\
\displaystyle\frac{\zeta(\alpha-2)}{2\zeta(\alpha-1)-\zeta(\alpha)} + \mathcal{O}(mr^{2-\alpha}) & , & mr^{1-\alpha}\ll mr^{2-\alpha}\ll 1
\end{array}
\right.\,.
\label{a>3}
\end{eqnarray}

From the results, (\ref{Deltas1<m_ll_r}) to (\ref{a>3}) above, one sees that the asymptotic behaviour of $\Delta_{s}^{(m)}(r)$ is one of the following form:
\begin{eqnarray}
\Delta_{s}^{(m)}(r) \sim \left\{
\begin{array}{l}
c_{1}r \\
 \\
\displaystyle\frac{r}{\ln r} \\
 \\
c_{2}r^{\omega} \\
 \\
c_{3}\ln r \\
 \\
c_{4}
\end{array}
\right.\,,
\label{allposs}
\end{eqnarray}
where $c_{1}$ ($0<c_{1}<1$), $c_{2}$, $c_{3}$, $c_{4}$ and $\omega$ do not depend on $r$. For instance, in the regime $1<m\ll r$ and for $1<\alpha<2$, one has $c_{2}=m/[C(m,\alpha)(2-\alpha)]$ and $\omega=2-\alpha$.

Each asymptotic form for $\Delta_{s}^{(m)}(r)$ leads to a different shortest path length $\ell$. The shortest path length between two randomly chosen vertices, $x_{a}$ and $x_{b}$, will be estimated to be of the order of the path length between $x_{a}$ and $0$, as stated in section \ref{shortest_path_length}. Symbolically, this implies
\begin{eqnarray}
\nonumber \sum_{i=1}^{\ell}\Delta_{s}^{(m)}(x_{i})\sim\textrm{distance from $x_{a}=x_{1}$ to the central node, which is $\mathcal{O}(t)$}\,,
\label{symb}
\end{eqnarray}
where $x_{2}$ is closer to the central node than $x_{1}=x_{a}$ by a distance $\Delta_{s}^{(m)}(x_{1})$ on average, $x_{3}$ is closer to the origin than $x_{2}$ by a distance $\Delta_{s}^{(m)}(x_{2})$ on average, and so on.

Since $x_{1}$ is at distance $\mathcal{O}(t)$ from the origin on average, it means that if $\Delta_{s}^{(m)}(r)=\mathcal{O}(1)$ (or $\Delta_{s}^{(m)}(r)\sim c_{4}$), then
\begin{eqnarray}
\ell(t)=\mathcal{O}(t)\,,\quad \textrm{when}\quad \Delta_{s}^{(m)}(r)\sim c_{4}\,.
\label{est_i}
\end{eqnarray}
On the other hand, suppose that $\Delta_{s}^{(m)}(r)\sim c_{1}r$, where $c_{1}\in(0,1)$ is a constant. In this case, one has the following sequence:
\begin{eqnarray}
\nonumber x_{2} & = & x_{1}-\Delta_{s}^{(m)}(x_{1}) = x_{1}-c_{1}x_{1} = \left(1-c_{1}\right)x_{1} \\
\nonumber x_{3} & = & x_{2}-c_{1}x_{2}=\left(1-c_{1}\right)x_{2} = \left(1-c_{1}\right)^{2}x_{1} \\
\nonumber x_{4} & = & x_{3}-c_{1}x_{3}=\left(1-c_{1}\right)^{3}x_{1} \\
\nonumber \cdots & & \\
x_{n} & = & \left(1-c_{1}\right)^{n-1}x_{1}
\label{seq}
\end{eqnarray}

From (\ref{seq}), one has
\begin{eqnarray}
n = 1+\frac{\ln x_{1}-\ln x_{n}}{\ln\left(\frac{1}{1-c_{1}}\right)}\,, \qquad c_{1}\in(0,1)\,.
\label{estimative}
\end{eqnarray}
Remembering that $x_{1}\sim\mathcal{O}(t)$ and if $n$ is of the order of $\ell$, then one has $\ell(t)\sim\ln t$. However, this kind of argument is suitable for $n$ not so small (since the estimation of $\Delta_{s}^{(m)}(r)$ is made for large values of $r$). Then, to estimate the shortest path length when $\Delta_{s}^{(m)}(r)\sim c_{1}r$ ($c_{1}\in(0,1)$), $n$ will be taken as being of order of $\ell$, which means that $x_{n}$ is close to the central node but \textit{not} the central node itself; in other words, $x_{n}\approx\mathcal{O}(1)$ and $n\approx\ell$. For large values of $t$, one has
\begin{eqnarray}
\ell(t)\sim\ln t\,,\quad \textrm{when}\quad \Delta_{s}^{(m)}(r)\sim c_{1}r \quad \textrm{for} \quad c_{1}\in(0,1)\,.
\label{est_ii}
\end{eqnarray}

For technical reason, the shortest path length for the other three cases will be estimated by using $\langle\Delta_{s}^{(m)}(r)\rangle := \sum_{u}\Delta_{s}^{(m)}(u)/t\approx\int^ {r}\Delta_{s}^{(m)}(u)du/t$ instead of $\Delta_{s}^{(m)}(r)$. From this approximation, one has
\begin{eqnarray}
\ell(t)\sim\ln t\,,\quad \textrm{when}\quad \Delta_{s}^{(m)}(r)\sim \frac{r}{\ln r}\,,
\label{est_iii}
\end{eqnarray}
\begin{eqnarray}
\ell(t)\sim t^{1-\omega}\,,\quad \textrm{when}\quad \Delta_{s}^{(m)}(r)\sim c_{2}r^{\omega}\,,
\label{est_iv}
\end{eqnarray}
and
\begin{eqnarray}
\ell(t)\sim\frac{t}{\ln t}\,,\quad \textrm{when}\quad \Delta_{s}^{(m)}(r)\sim \ln r\,.
\label{est_v}
\end{eqnarray}



\section*{References}

\begin{thebibliography}{999}

\bibitem{B85} Bollob\'as B, \textit{Random Graphs} (Academic Press, London, 1985)

\bibitem{WS98} Watts D J, Strogatz S H, 1998 \textit{Nature} \textbf{393} 440

\bibitem{NW99} Newman M E J, Watts D J, 1999 \textit{Phys. Rev. E} \textbf{60} 7332

\bibitem{BA99} Barab\'asi A L, Albert R, 1999 \textit{Science} \textbf{286} 509

\bibitem{VB85} Viana L, Bray A J, 1985 \textit{J. Phys. C} \textbf{18} 3037

\bibitem{MM04} Marinari E, Monasson R, 2004 \textit{J. Stat. Mech.} P09004

\bibitem{ZM06} Zdeborov\'a L, M\'ezard M, 2006 \textit{J. Stat. Mech.} P05003

\bibitem{HM08} Hase M O, Mendes J F F, 2008 \textit{J. Phys. A} \textbf{41} 145002

\bibitem{K00} Kleinberg J, 2000 \textit{Nature} \textbf{406} 845 -- see also available articles on http://www.cs.cornell.edu/home/kleinber

\bibitem{MvN04} Martel C, Nguyen V K, in Proc. of the 23rd ACM Symposium on Principles of Distributed Computing PODC'2004

\bibitem{BBCdSK04} Berger N, Borgs C, Chayes J T, D'Souza R M, Kleinberg R D, \textit{Lecture Notes in Computer Science} \textbf{3142}, 208 (2004)

\bibitem{FKP02} Fabrikant A, Koutsoupias E, Papadimitriou C H, \textit{Lecture Notes in Computer Science} \textbf{2380}, 781 (2002)

\bibitem{SM03} Sen P, Manna P, 2003 \textit{Phys. Rev. E} \textbf{68} 026104

\bibitem{RbA06} Roberson M R, ben-Avraham D, 2006 \textit{Phys. Rev. E} \textbf{74} 017101

\bibitem{YJB02} Yook S H, Jeong H, Barab\'asi A L, 2002 \textit{Proc. Natl. Acad. Sci.} \textbf{99}, 13382

\bibitem{MS02} Manna S S, Sen P, 2002 \textit{Phys. Rev. E} \textbf{E}, 066114

\bibitem{XBS02} Xulvoi-Brunet R, Sokolov I M, 2002 \textit{Phys. Rev. E} \textbf{66}, 026118

\bibitem{B03} Barth\'elemy M, 2003 \textit{Europhys. Lett.} \textbf{63}, 915

\bibitem{MMK05} Masuda N, Miwa H, Konno N, 2005 \textit{Phys. Rev. E} \textbf{71}, 036108

\bibitem{BBV05} Barrat A, Barth\'elemy M, Vespignani A, 2005, \textit{J. Stat. Mech.} P05003

\bibitem{W88} Waxman B M, 1988 \textit{IEEE J. Select. Areas Commun.} \textbf{6}, 1617

\bibitem{H75} Hardy G H, Wright E M, \textit{An Introduction to the Theory of Numbers} (Clarendon Press, Oxford, 1975) -- page 7

\bibitem{DM00} Dorogovtsev S N, Mendes J F F, 2000 \textit{Phys. Rev. E} \textbf{62} 1842

\bibitem{DM01} Dorogovtsev S N, Mendes J F F, 2001 \textit{Phys. Rev. E} \textbf{63} 056125

\bibitem{DM03} Dorogovtsev S N, Mendes J F F, \textit{Evolution of Networks} (Oxford University Press, Oxford, 2003) -- Chapter 5

\end{thebibliography}
\end{document}